\documentstyle[aps,prl,epsfig,multicol]{revtex}

\begin{document}
\title{Effect of one-ion $L$-$S$ coupling on magnetic properties of a correlated
spin-orbit system}
\author{Zu-Jian Ying$^{1,2,3}$, Xi-Wen Guan$^4$, Angela Foerster$^1$, Itzhak Roditi$%
^3$, Bin Chen$^2$}
\address{1. Instituto de F\'\i sica da UFRGS, Av. Bento Gon\c calves, 9500, Porto
Alegre, 91501-970, Brasil\\
2. Hangzhou Teachers College, Hangzhou 310012, China\\
3. Centro Brasileiro de Pesquisas F\'\i sicas, Rua Dr. Xavier Sigaud 150,
22290-180 Rio de Janeiro, RJ, Brasil\\
4. Department of Theoretical Physics, Research School of Physical Sciences
and Engineering, and Centre for Mathematics and its Applications,
Mathematical Sciences Institute, Australian National University, Canberra
ACT 0200, Australia}
\date{ }
\maketitle

\begin{abstract}
By introducing a basis for a novel realization of the SU(4) Lie algebra, we
exactly solve a spin-orbital chain with one-ion $L$-$S$ coupling (OILSC) via
the Bethe ansatz (BA) approach. In the context of different Land\'e $g$
factors of the spin and orbital sectors, the OILSC results in rich and novel
quantum phase transitions. Some accurate analytical expressions for the
critical fields are obtained. Both spin ordering and orbital ordering are
found in a gapped singlet phase. The system exhibits many interesting
phenomena such as nonvanishing magnetization of the singlet phase,
multi-entrance of the singlet in the ground state when the field varies,
unsymmetric magnetization in two-component phase and magnetization crossover
for different OILSC.
\end{abstract}


\begin{multicols}{2}

Strongly-correlated electron systems in the presence of orbital degeneracy
have attracted much interest due to experimental advances in
transition-metal compounds\cite{Nagaosa}, and the orbital ordering as well
as the orbital density wave has been observed in a family of manganites\cite
{Saitoh}. Among the experimental findings are materials related to
spin-orbital systems in one dimension, such as tetrahis (dimethylamino)
ethylene-C$_{60}$\cite{Arovas}, artificial quantum dot arrays\cite{Onufriev}%
, Na$_2$Ti$_2$Sb$_2$O and NaV$_2$O$_5$ degenerate chains\cite{Pati}, and so
forth. A well-known spin-orbital model is the SU(4) model\cite{LiSU4} which
is exactly solvable in one dimension\cite{LiSU4BA,Sutherland}. The
competition of orbital degree of freedom with the spin results in three
quantum phase transitions (QPT's) in a magnetic field, according to both
numerical\cite{GuSU4} and analytic\cite{Ying} analysis. Deviations from the
SU(4) symmetry can be caused by variations of different-site interaction\cite
{Arovas,Pati,Mila,Azaria}. Another possibility for deviation was considered
for an SU(2)$\otimes $SU(2) Ising-type same-site anisotropy in Ref.\cite
{Gu,Ying}, which may occur when the external field turns the spin oriented
in parallel to the orbital angular momentum frozen in some direction by a
crystalline field. A more generally-existing same-site interaction is SU(3)$%
\otimes $U(1) one-ion $L$-$S$ coupling (OILSC), the strength of which varies
substantially in a range of 0$\sim $10$^4$cm$^{-1}$(1cm$^{-1}\sim $1K$\sim $%
10$^{-4}$eV) for different elements\cite{Cowan}. As it is well understood
now, $L$-$S$ coupling can cause magnetic anisotropy\cite{Slichter}, as far
as a single ion is concerned. An interesting problem is to investigate the
influence of such an OILSC on the collective properties, especially the
QPT's of strongly-correlated spin-orbital systems in a magnetic field. It
can be expected that the OILSC will bring about novel physics in the
competition of spin and orbital degrees of freedom.

In the present paper we shall exactly solve a spin-orbital chain with such
an OILSC via the Bethe ansatz (BA) approach. In the context of the different 
$g$ factors of spin and orbit sectors, the OILSC leads to a rich variety of
novel QPT's, with the analytic critical fields explicitly obtained. Both
spin ordering and orbital ordering exist in a gapped singlet phase,
accompanied by a nonvanishing magnetization. The energy of the singlet is no
longer invariant in the field due to the different $g$ factors, which brings
about five consecutive QPT's. It also contributes to unsymmetric changes of
the total magnetization within a same phase in weak and in strong fields.

We consider an $N$-site chain with the Hamiltonian 
\begin{eqnarray}
{\cal H} &=&{\cal H}_0+{\cal H}_{L-S}+{\cal M},\ {\cal H}_0=%
\sum_{i=1}^NP_{i,i+1},  \nonumber \\
{\cal H}_{L-S} &=&\lambda \sum_i{\vec l}_i\cdot {\vec s}_i,\ {\cal M}%
=-g_sH\sum_i(g_ss_i^z+g_tl_i^z),
\end{eqnarray}
where $\vec s$ and ${\vec l}$ are spin-1/2 operators for spin and orbit and $%
g_s$ and $g_t$ are the corresponding Land\'e $g$ factors, $P_{i,i+1}=(2\vec s%
_i\cdot \vec s_{i+1}+\frac 12)(2{\vec l}_i\cdot {\vec l}_{i+1}+\frac 12)$
exhanges the neigher-site states. $H$ is the magnetic field and periodic
boundary conditions are applied throughout. ${\cal H}_{L\text{-}S}$ is the
OILSC from the $N$ sites and note that electrons have positive $\lambda $
whereas holes have negative $\lambda $\cite{Slichter}. In the absence of the
field, ${\cal H}_{L\text{-}S}$ breaks the system symmetry into SU(3)$\otimes 
$U(1) when the bulk part ${\cal H}_0$ is SU(4) invariant\cite{LiSU4}. The
basis consists of singlet and triplet of the SU(2) Lie algebra \{$%
s^{+}+l^{+},s^{-}+l^{-},s^z+l^z\}$, which is valid in the presence of the
field only if $g_s=g_t$ as in the integrable spin ladder\cite{WangY,Xiwen}.
However, ${\cal H}_{L\text{-}S}$ and ${\cal M}$ do not commute due to the
different $g$ factors of spin and orbit, the conventional singlet and
triplet are no longer their common eigenstates and fail to be the solution
in the spin-orbital system. Nevertheless, we can diagonalize ${\cal H}_{L%
\text{-}S}+{\cal M}$ as a whole, which requires a new basis for all sites 
\[
\varphi _0=\frac{\phi _1-y^{-1}\phi _2}{\sqrt{1+y^{-2}}},\ \varphi _1=\phi
_3,\ \varphi _2=\frac{\phi _1+y\phi _2}{\sqrt{1+y^2}},\ \varphi _3=\phi _4, 
\]
where $\phi _1=\left| \uparrow \downarrow \right\rangle $, $\phi _2=\left|
\downarrow \uparrow \right\rangle $, $\phi _3=\left| \uparrow \uparrow
\right\rangle $, $\phi _4=\left| \downarrow \downarrow \right\rangle $
denote site state $\left| s^zl^z\right\rangle $ and $y^{\pm 1}=\pm
g_{-}H/\lambda +\gamma (H/\lambda )$, $\gamma (h)=(1+g_{-}^2h^2)^{1/2}$, $%
g_{\pm }=g_s\pm g_t$. The new realization of the SU(4) Lie algebra takes $%
S_{mn}\varphi _i=\delta _{ni}\varphi _m$, 
\begin{eqnarray}
S_{10} &=&p(y)r(y)[yl^{+}(1/2+s^z)-s^{+}(l^z+1/2)],  \nonumber \\
S_{30} &=&p(y)r(y)[ys^{-}(1/2-l^z)-l^{-}(1/2-s^z)],  \nonumber \\
S_{12} &=&r(y)[l^{+}(1/2+s^z)+ys^{+}(1/2+l^z)],  \nonumber \\
S_{32} &=&r(y)[s^{-}(1/2-l^z)+yl^{-}(1/2-s^z)],\ S_{13}=s^{+}l^{+}, 
\nonumber \\
S_{02} &=&p(y)r^2(y)[y^2s^{+}l^{-}-s^{-}l^{+}+y(s^z-l^z)],  \label{SU4}
\end{eqnarray}
where $r(y)=1/\sqrt{1+y^2}$, $p(y)=y/\left| y\right| $, other operators are
their conjugates $S_{nm}=S_{mn}^{\dagger }$. For a certain choice of the
basis order, the three Cartan operators $I_k$ ($k=1,2,3)$ can be generated
by commutations, e.g., $I_1=[S_{23},S_{32}]$, $I_2=[S_{12},S_{21}]$ and $%
I_3=[S_{01},S_{10}]$ for basis order $(\varphi _0$,$\varphi _1$,$\varphi _2$,%
$\varphi _3)^T$. The new basis \{$\varphi _0\}$ and \{$\varphi _1$,$\varphi
_2$,$\varphi _3\}$ form the singlet and triplet of a new SU(2) Lie algebra \{%
$A^{+},A^{-},A^z$\} where $A^z=s^z+l^z$ and 
\[
A^{\pm }=\sqrt{2}r(y)[(y-1)\left( s^{\pm }l^z-l^{\pm }s^z\right) +\frac{y+1}2%
(s^{\pm }+l^{\pm })], 
\]
though the field breaks the symmetry into four U(1)'s.

We notice that a time-dependent single $L$-$S$ coupled particle was solved
by an SU(2) gauge transformation\cite{SJWang}, but here we consider a
many-body system with $N$ correlated ions. Note that $P_{i,j}$ is still the
permutation operator in the new basis, and ${\cal H}_{L\text{-}S}+{\cal M}$
commutes with the SU(4) bulk part ${\cal H}_0$. Based on the above new SU(4)
realization we can exactly solve the model via the BA approach\cite
{Sutherland} with the BA equations and the eigenenergy 
\begin{eqnarray*}
-\prod_{m=1}^{M^{(k)}}\Xi _1(\mu _{j,m}^{k,k}) &=&\prod_{m=1}^{M^{(k+1)}}\Xi
_{\frac 12}(\mu _{j,m}^{k,k+1})\prod_{m=1}^{M^{(k-1)}}\Xi _{\frac 12}(\mu
_{j,m}^{k,k-1}), \\
E &=&-\sum_{j=1}^{M^{(1)}}2\pi a_1(\mu _j^{(1)})+\sum_{i=1}^4E_iN_i,
\end{eqnarray*}
where $\Xi _x(\mu _{j,m}^{k,l})=(\mu _j^{(k)}-\mu _m^{(l)}-xi)/(\mu
_j^{(k)}-\mu _m^{(l)}+xi)$, $\mu _j^{(0)}=0,M^{(0)}=N,M^{(4)}=0,$ and $1\leq
k\leq 3;$ $a_n(\mu )=\frac 1{2\pi }\frac n{\mu ^2+n^2/4}$, $E_0=-\frac 
\lambda 4-\frac \lambda 2\gamma (H/\lambda )$, $E_1=\frac \lambda 4-\frac 12%
g_{+}H$, $E_3=\frac \lambda 4+\frac 12g_{+}H$ and $E_3=-\frac \lambda 4+%
\frac \lambda 2\gamma (H/\lambda )$. $N_i$ is the total site number in state 
$\varphi _i$. The basis order is chosen as ($\varphi _{P_1}\varphi
_{P_2}\varphi _{P_3}\varphi _{P_4}$)$^T$, where $P_i\in \{1,2,3,4\}$ and $%
\varphi _{P_1}$ is energetically the most favorable while $\varphi _{P_4}$
is the least favorable. The relation $N_{P_i}=M^{(i-1)}-M^{(i)}$ gives
another form of energy $E=\sum_{j=1}^{M^{(1)}}g^{(1)}(\mu
_j)+g^{(2)}M^{(2)}+g^{(3)}M^{(3)}$. The field shifts the order from ($%
\varphi _0$,$\varphi _1$,$\varphi _2$,$\varphi _3$)$^T$ to ($\varphi _1$,$%
\varphi _0$,$\varphi _2$,$\varphi _3$)$^T$ for $\lambda >0$, and ($\varphi
_1 $,$\varphi _2$,$\varphi _3$,$\varphi _0$)$^T$ to ($\varphi _1$,$\varphi
_2 $,$\varphi _0$,$\varphi _3$)$^T$ for $\lambda <0$, at point $%
H_R=g_{+}\left| \lambda \right| /(2g_sg_t)$. Following Ref.\cite{string} and
Ref.\cite{Y-Y} one can obtain the ground state (GS) equations for three
dressed energy $\epsilon ^{(i)}$, 
\begin{equation}
\epsilon ^{(i)}=g^{(i)}-a_2*\epsilon ^{(i)-}+a_1*(\epsilon
^{(i-1)-}+\epsilon ^{(i+1)-}),  \label{TBA}
\end{equation}
where $\epsilon ^{(0)}=\epsilon ^{(4)}=0$ and the symbol $*$ denotes the
convolution. Fermi seas filled by negative $\epsilon ^{(i)-}$ form the GS.

In the absence of the field, the GS is composed of all the four basis
components when $\lambda _{c-}<\lambda <\lambda _{c+}$ where $\lambda
_{c+}=4 $, $\lambda _{c-}=-2a_0/3$ and $a_0=(\sqrt{3}\pi -3\ln 3)/2$. But a
strong positive OILSC $\lambda >\lambda _{c+}$ keeps the triplet gapful, the
GS only consists of singlet and the system exhibits spin and orbital
ordering, as shown later on. The first QPT will occur when the field brings
down the triplet $\varphi _1$ and closes the gap $\Delta =g^{(1)}(0)$ at a
critical field 
\begin{equation}
H_{c0}=\frac{\lambda -8+\sqrt{\lambda ^2-\Delta _g^216(\lambda -4)}}{%
g_{+}(1-\Delta _g^2)}  \label{Hc0}
\end{equation}
where $\Delta _g=g_{-}/g_{+}$. Further increase of the field will exhaust
all the singlet and so fully polarize the GS at 
\begin{equation}
H_f=\frac{\lambda +8+\sqrt{\lambda ^2+\Delta _g^216(\lambda +4)}}{%
g_{+}(1-\Delta _g^2)},  \label{Hf}
\end{equation}
which is valid for all values of $\lambda $. For a weak OILSC $0<\lambda
<\lambda _{c+}$, the field brings out the components $\varphi _3,\varphi
_2,\varphi _0$ in turn at three different critical points.

For $\lambda <0$, the single-ion energy $E_0$ of the singlet $\varphi _0$
rises when the field becomes stronger. The triplet component $\varphi _3$
has lower energy in zero field but will be raised by the field in a quicker
way than $\varphi _0$. The energy competition of $\varphi _3$ and $\varphi _0
$ results in new kinds of QPT's. In the four-component GS a weak negative
OILSC does not expel the singlet far away from the triplet. In an increasing
field $\varphi _3$ rises in energy more quickly and soon goes beyond $%
\varphi _0$, so $\varphi _3$ gets out of the GS before $\varphi _0$. There
are three critical fields consecutively exhausting $\varphi _3,\varphi _0$
and $\varphi _2$ from the GS. However, a stronger negative OILSC leaves a
longer distance between the triplet and singlet. Before $\varphi _3$ covers
this distance to $\varphi _0$, the singlet $\varphi _0$ has already risen
completely out of the GS, which happens in the first critical point. But $%
\varphi _3$ soon draws very near to $\varphi _0$ such that $\varphi _0$
becomes close enough to the GS, in which $\varphi _3$ still lies, and is
drawn back into the GS. This brings about the second QPT. Afterwards, the
field overwhelms the influence of OILSC and pump out $\varphi _3,\varphi _0$
and $\varphi _2$ one by one, which gives other three QPT's. Therefore the
system undergoes five consecutive QPT's (5-QPT) in this case, the variations
of the components in the GS are: 1230$\rightarrow $123$\rightarrow $1230
(and 1203)$\rightarrow $120$\rightarrow $12$\rightarrow $1, where we denote
the component $\varphi _i$ by number $i$ and each arrow indicates a QPT.
Such unusual 5-QPT occur near $\lambda _{c-}$ ($\lambda _{5c}<\lambda
<\lambda _{c-}$, $\lambda _{5c}\doteq -0.712$, for $g_s=2.0$, $g_t=1.0$).
When the OILSC is negatively stronger than $J_{c-}$, the singlet is gapful
and does not exist in the GS before the field is applied. But the quicker
rising of $\varphi _3$ will get the singlet involved in the GS, as what
happens in the second QPT of the 5-QPT. In such case, four consecutive QPT's
(4-QPT) will be observed: 123$\rightarrow $1230 (and 1203)$\rightarrow $120$%
\rightarrow $12$\rightarrow $1. A negatively stronger OLSC than $\lambda _Q$
will expel the singlet too far away from the GS composed of the triplet. The
triplet component $\varphi _3$ itself will get out of the GS before it
becomes adhesive and draw the singlet into the GS. We then have only the two
QPT's.

\begin{figure}[t]
\setlength\epsfxsize{75mm}
\epsfbox{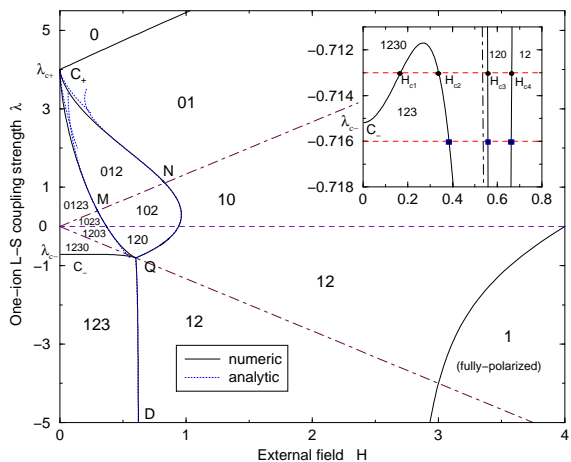}
\caption{Phase diagram of quantum phase transitions (QPT's) in magnetic field, $%
g_s=2.0$, $g_t=1.0$. The number i labels the state $\varphi _i$, e.g., the
state components in phase 0123 are $\varphi _0\varphi _1\varphi _2\varphi _3$
in which the singlet $\varphi _0$ is energetically the most favorable
whereas the triplet component $\varphi _3$ is the least favorable. The
discrepancy between the analytic and numerical critical fields is not
visible for most region of C$_{+}$NQD and C$_{+}$MQ. The boundary of singlet
phase and the full-polarized point are exact. INSET: the five consecutive
QPT's (dark dots) and four QPT's (dark boxes), their full-polarized points
are relatively far away.
}
\label{phase}
\end{figure}

\begin{figure}[t]
\setlength\epsfxsize{75mm}
\epsfbox{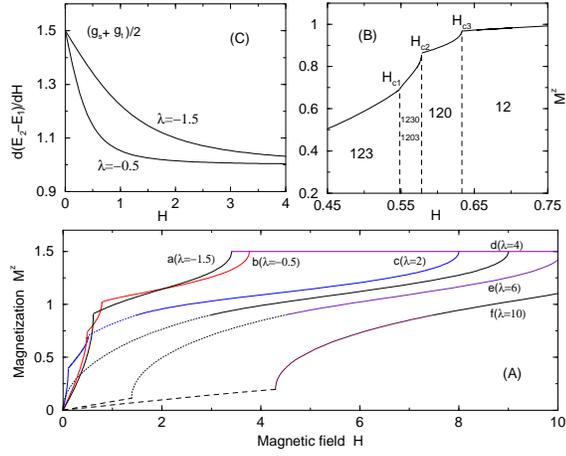}
\caption{(A) Typical magnetization behaviors in different QPT's, $g_s=2.0$, $%
g_t=1.0$. The dashed lines are gapped singlet phase with a nonzero
magnetization $M^z$, both spin and orbital ordering can be observed in this
phase. Curve d provides a full view of unsymmetric growth of $M^z$, the
dotted-line region (phase 01) in weak field climbs faster than the
solid-line region (phase 10) in strong field. Also note the crossover
between curve a and b before full polarization. (B) Magnetization of four
consecutive QPT's, $\lambda =-0.76,$ $H_{c4}=3.65$, the numbers denote the
phases as in Fig.\ref{phase}. (C) Comparison on increasing speed of energy
difference $E_2-E_1$ in the field for $\lambda =-1.5$ and $\lambda =-0.5$,
the former always has a larger speed.
}
\label{Mz}
\end{figure}

We present a detailed phase diagram in Fig.\ref{phase}, the inset
illustrates the 5- and 4-QPT. 5-QPT exist if C$_{-}$Q tends to rise up at C$%
_{-}$, with requirement on the g factors $g_t<(1-\Delta _{g5})g_s/(1+\Delta
_{g5})\doteq 0.516g_s$, where $\Delta _{g5}\doteq 3[(1+w)a_0]^{1/2}/(4\pi )$
and $w=2/3$, satisfied for most typical spin-orbital systems. When the
triplet $\varphi _3$ attracts the singlet $\varphi _0$ into the GS, some
density of $\varphi _3$ themselves are attracted upwards a bit by $\varphi
_0 $, which causes a quicker increase in magnetization. This
attraction-and-counter-attraction is more sensitive near the point Q where $%
\varphi _3$ has small proportion in the GS. A 4-QPT magnetization near point
Q is plotted in Fig.\ref{Mz}B. The magnetization in the 5-QPT case has less
apparent changes in the first two transition points due to heavier density
of $\varphi _3$.

Besides the exact singlet phase critical point (\ref{Hc0}) and
full-polarized point (\ref{Hf}), other critical points of the QPT can be
obtained from Wiener-Hopf method\cite{WH} 
\begin{eqnarray*}
&&H_c^{C_{+}NQ}\doteq H_{+}+\tau k_{+}G_{-1,1,-1}^2(\lambda ,H_{+}), \\
&&H_c^{QD}\doteq H_{-}+\tau k_{-}G_{-1,1,-1}^2(\lambda ,H_{-}), \\
&&H_c^{C_{+}M}\doteq H_0+\tau k_0[wG_{\frac 12,-\frac 12,\frac 32}^2(\lambda
,H_0)+G_{-\frac 12,\frac 12,\frac 32}^2(\lambda ,H_0)], \\
&&H_c^{MQ}\doteq H_0+\tau k_0[wG_{-1,1,0}^2(\lambda ,H_0)+G_{-\frac 12,\frac 
12,\frac 32}^2(\lambda ,H_0)],
\end{eqnarray*}
where $k_0=3g_{+}^{-1}$, $\tau =1/(4\pi ^2)$, $H_0=(a_0-\lambda
/2)g_{+}^{-1},$ 
\begin{eqnarray*}
&&H_{\pm }=\frac{\lambda _{\pm }-3^{\pm 1}\sqrt{(\lambda _{\pm }^2-3^{1\pm
1}\lambda ^2)\Delta _g^2+\lambda ^2}}{3^{(1\mp 1)/2}g_{+}(1-9^{\pm 1}\Delta
_g^2)}, \\
&&G_{l,m,n}(\lambda ,H)=l\lambda /2+mg_{+}H/2+n\sqrt{\lambda ^2+g_{-}^2H^2}%
/2, \\
&&k_{\pm }=\frac 4{3^{(1\mp 1)/2}g_{+}}(1+\frac{3^{(1\pm 3)/2}\Delta
_g^2g_{+}H_{\pm }}{\lambda _{\pm }-3^{(1\mp 1)/2}g_{+}H_{\pm }})^{-1},
\end{eqnarray*}
and $\lambda _{\pm }=8\ln 2\pm \lambda $. $H_{\pm }$ and $H_0$ are the
contributions of infinity Fermi points (FP), the other terms come from the
revision of the finite FP. For $g_s=2g_t$, $H_{+}$ can be simplified as $%
H_{+}=[\lambda _{+}^2-9\lambda ^2]/[2g_{+}\lambda _{+}]$. The above analytic
results are compared with the numerical ones in Fig.\ref{phase}, which shows
an excellent accuracy. For C$_{+}$NQD and C$_{+}$MQ, deviations only occur
near C$_{+}$ due to small FQ invalid for Wiener-Hopf method, but an
expansion based on small FP can be carried out and $H_c\cong (2c/\pi
)g_{+}^{-1}(\lambda _{c+}-\lambda )^{3/2}$, $c=1$ for C$_{+}$N and $c=(\ln
2)/3$ for C$_{+}$M. Setting $\lambda =0$ in $H_c^{MQ}$, $H_c^{C_{+}NQ}$ and $%
H_f$ will recover the very accurate expressions of the SU(4) model obtained
in Ref.\cite{Ying}.

A remarkable progress in experiments of spin-orbital systems is the
observation of orbital ordering in some transition-metal compounds\cite
{Saitoh}, which can be also found in the present model. The expectations of
the spin ${\vec s}$ and orbital angular momentum ${\vec l}$, the
magnetization $M^z=g_ss^z+g_tl^z$ can be obtained for a singlet site 
\[
\langle {\vec s}\rangle _0=-\langle {\vec l}\rangle _0=g_{-}H[2\lambda
\gamma (H/\lambda )]^{-1},\ \langle M^z\rangle _0=g_{-}\langle {\vec s}%
\rangle _0. 
\]
Many transition-metal elements involve a strong OILSC and the GS of the
system can be located in the singlet phase, in which all sites are occupied
by the singlet. Consequently, the spins will be aligned ferromagnetically in
the $z$ direction while all the orbitals are ferromagnetically ordered in
the opposite direction. Therefore, both orbital ordering and spin ordering
can be observed. In addition, unlike the conventional singlet, the
magnetization $\langle M^z\rangle _0$ is no longer zero in the field, as
shown by the dash lines of curves {\it e} and {\it f} in Fig.\ref{Mz}A. $%
\langle M^z\rangle _0$ increases almost in a linear way due to a small value
of $H/\lambda $ before the singlet phase is transited to another phase at $%
H_{c0}$. But the increasing rate $d\langle M^z\rangle _0/dH$ becomes smaller
when the field gets stronger. This will lead to unsymmetric increasing of
the magnetization in weak and in strong fields. The dotted regions of curves 
{\it c}, {\it d} and {\it e} in Fig.\ref{Mz}A show a quicker climbing than
the regions thereafter, which happens in a same phase with only two
components $\varphi _0$ and $\varphi _1$ in the GS. Curve {\it d} only
involves these two components from the beginning and provides a full view of
this unsymmetry. Besides the slowdown of $d\langle M^z\rangle _0/dH$,
another contribution to the unsymmetry comes from the variation in the
energy-difference increasing rate. The energy difference $\left|
E_1-E_0\right| $ between the GS components $\varphi _1$ and $\varphi _0$
varies in a slower rate $d\left| E_1-E_0\right| /dH$ when the field reaches
higher.

We also present two typical magnetization for negative OILSC, shown by
curves {\it a} and {\it b} in Fig.\ref{Mz}A. The corresponding QPT's are: (%
{\it a}) 123$\rightarrow $12$\rightarrow $1; ({\it b}) 1230 (and 1203)$%
\rightarrow $120$\rightarrow $12$\rightarrow $1. One can notice the
crossover of the two curves in phase 12 before full polarization, where only
two components $\varphi _1$ and $\varphi _2$ are left in the GS. Curve {\it b%
} lies in four-component phase 1230 in the absence of the field, while a
stronger OILSC expels the singlet off the GS and curve {\it a} starts from
triplet phase 123. The negative $\lambda $ makes $\varphi _0$ rising in
energy level while $\varphi _2$ sinks when the field is applied. But the
lowest state $\varphi _1$ sinks fast and so all the other components are
moving out of the GS. Thus there is one more component $\varphi _0$ getting
out of the GS for case {\it b} than case {\it a}. Therefore, the proportion
of the most energetically favorable component $\varphi _1$, with a positive
magnetization $g_{+}/2$, increases more quickly in case {\it b} than in case 
{\it a}, which gives a quicker rise of magnetization in curve {\it b} before
the first QPT. The component $\varphi _3$ in case {\it b} has smaller
proportion in the beginning of a four-component phase 1230, while it has
larger proportion in case {\it a} starting from a three-component phase 123.
Consequently, $\varphi _3$ gets out of the GS earlier in case {\it b} and
the first QPT occurs ahead of case {\it a}. But the rise of curve {\it b}
soon slows down after its earlier first QPT, since the $\varphi _0$ has
weaker negative magnetization than $\varphi _3$ in case {\it a}. After $%
\varphi _3$ gets out in case {\it a}, curve {\it b} soon rises beyond curve 
{\it a} again as it has two components $\varphi _2$ and $\varphi _0$ moving
out while only one component $\varphi _2$ in case {\it a}. As a result, case 
{\it b} has higher magnetization than case {\it a} when they both come to
phase 12. But in this phase, $\varphi _2$ rises away from the lower state $%
\varphi _1$ always more quickly in case {\it a} with a stronger $\lambda $,
as shown by $d\left| E_2-E_1\right| /dH$ in Fig.\ref{Mz}B. Therefore, $%
\varphi _2$ gets out faster in case {\it a}, and its lower magnetization
rises in advance of case {\it b} and reaches the saturation point first.
This results in the crossover.

Finally we stress only in the context of the different $g$ factors of spin
and orbit that the OILSC leads to these properties. If $g_s=g_t$\cite{Xiwen}, 
the novel
phenomena including the five consecutive QPT's, the spin and orbtal
ordering, nonzero magnetization of the singlet, the unsymmetric
magnetization and the magnetization crossover, will all disappear.

We thank Huan-Qiang Zhou, You-Quan Li and Murray T. Batchelor for helpful
discussions. ZJY thanks FAPERGS and FAPERJ for support. XWG thanks
Australian Research Council. AF thanks FAPERGS and CNPq. IR thanks PRONEX
and CNPq. BC thanks Zhejiang Natural Science Foundation RC02068.

\end{multicols}



\end{document}